\documentclass[sigconf,nonacm]{acmart}    %

\usepackage{color}
\usepackage{bm}
\usepackage{subcaption}
\usepackage{multirow}
\usepackage{enumitem}

\providecommand{\added}[2][]{{#2}}
\providecommand{\deleted}[2][]{}

\newif\ifshowchanges
\showchangesfalse %

\definecolor{forestgreen}{RGB}{0, 140, 0}
\definecolor{cerulean}{RGB}{42, 82, 190}
\definecolor{blue}{RGB}{0, 0, 220}

\definecolor{changecolor}{RGB}{20, 50, 200}

\definecolor{abecolor}{RGB}{0, 140, 0}
\definecolor{abe_addcolor}{rgb}{0,0.3,0.8}
\definecolor{abe_removecolor}{rgb}{0.9,0.2,0.1}

\definecolor{nhancolor}{RGB}{0, 100, 50}

\ifshowchanges

    \newcommand{\adelete}[1]{{\leavevmode\textcolor{abe_removecolor}{{DeletedPart}}}}
    
    \newcommand{\acomment}[1]{\comment[id={\AbeD}]{#1}}
    \newcommand{\changenote}[2][]{\added[comment={#2}]{}}

    \newenvironment{revised}{
    \color{changecolor}
    }{
    \color{black}
    }

    \newenvironment{nhan_revised}{
    \color{nhancolor}
    }{
    \color{black}
    }

\else

    \newcommand{\adelete}[1]{}
    
    \newcommand{\acomment}[1]{}
    \newcommand{\changenote}[2][]{}

\fi

\newif\ifsubmit
\submittrue

\input{macros/NotesAndEditsMacros}

\newcommand{\beq}{\begin{equation}}
\newcommand{\eeq}{\end{equation}}

\newcommand{\baeq}{\beq\begin{aligned}}
\newcommand{\eaeq}{\end{aligned}\eeq}

\newcommand{\defineterm}[1]{\emph{#1}}

\newcommand{\namedparagraph}[1]{\vspace{0.2cm}\noindent\textbf{#1:}}

\newcommand{\workflowname}{Hybrid Cinematography}
\newcommand{\Workflowname}{\workflowname{}}

\newcommand{\burdenterm}{hallucination risk}
\newcommand{\Burdenterm}{Hallucination risk}
\newcommand{\BurdenTerm}{Hallucination Risk}

\newcounter{formativefindingcounter}
\newcounter{designgoalcounter}
\newcommand{\capturedmove}{captured move}

\newcommand{\targetmove}{target move}

\newcommand{\proxy}{proxy}
\newcommand{\previs}{previs}

\newcommand{\lam}{\lambda}
\newcommand{\camcap}{C_{\mathrm{c}}}      %
\newcommand{\camtar}{C_{\mathrm{o}}}      %
\newcommand{\camlam}{C_{\lam}}            %

\makeatletter
\DeclareRobustCommand{\projectlink}{%
  \if@ACM@anonymous\else
    \newline Video and examples: {\color{ACMBlue}\url{https://hybridcinematography.github.io/}}%
  \fi}
\makeatother

\begin{document}

\newcommand{\PaperTitle}{\workflowname{}: Previsualizing and Managing Hallucination Risk in Generative Video Reshooting}

\newcommand{\ShortPaperTitle}{\workflowname{}}

\title[\ShortPaperTitle{}]{\PaperTitle{}}

\author{Nhan Tran}
\affiliation{%
  \institution{Cornell University}
  \city{Ithaca, NY}%
  \country{USA}}
\email{nhan@cs.cornell.edu}

\author{Neal Wadhwa}
\affiliation{%
  \institution{Google}
  \city{New York, NY}%
  \country{USA}}
\email{nealw@google.com}

\author{Abe Davis}
\affiliation{%
  \institution{Cornell University}
  \city{Ithaca, NY}%
  \country{USA}}
\email{abedavis@cornell.edu}

\author{Stefan Stojanov}
\affiliation{%
  \institution{Google}
  \city{New York, NY}%
  \country{USA}}
\email{stojanov@google.com}

\begin{abstract}

On a film set, the camera move is committed during a take. Generative video reshooting lets filmmakers change it afterward, but may require hallucinating unrecorded content, a gap sometimes discovered only after leaving the set. We present \workflowname{}, a workflow that bridges physical capture and generative reshooting to manage \burdenterm{} while filmmakers can still act on it. Using an editable 3D shot plan and a proxy of the take, our previsualization evaluates \burdenterm{} in real time. Seeing where the take lacks support, filmmakers can iteratively adjust the plan, explore moves that balance capture and generation, shoot guided pickups, or knowingly accept hallucination. We demonstrate the workflow through a mobile augmented reality application for on-set planning, capture, and review, and an offline pipeline for existing video. A study with experienced filmmakers reveals how previsualizing risk informs camera decisions and exposes tensions between creative intent and generative hallucination.

\end{abstract}



\begin{teaserfigure}
  \includegraphics[width=\textwidth]{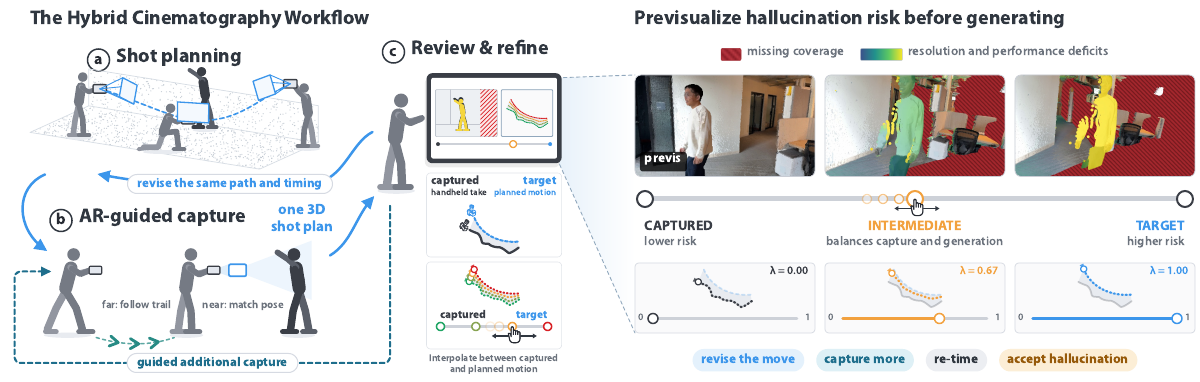}
  \caption{\nt{We present \workflowname{}, a workflow connecting capture and generative post-production, built around one editable 3D shot plan that every stage reads (left). We demonstrate the workflow in a mobile augmented reality (AR) app. (a) Before or after capture, filmmakers plan the \targetmove{} as keyframes by walking the scene, then set its timing. (b) In capture, the same plan drives the AR guidance. (c) In review, a previsualization (\previs{}) checks the recorded take against that plan and marks what the model would have to invent, its \defineterm{\burdenterm{}}. Every move between the \capturedmove{} and the \targetmove{} is a possible reshoot, asking the model to invent what the take does not support. Filmmakers navigate that range with a slider and see the risk of each move before anything is generated (right). They can then revise the move, capture more, re-time, or knowingly accept hallucination, while still on set.}\projectlink}
  \label{fig:teaser}
  \Description{Two-part figure. The left part, titled The Hybrid Cinematography Workflow, shows three stages connected in a loop around one 3D shot plan. In (a) Shot planning, people walk a scene and place camera keyframes in AR, joined by a dashed blue path. In (b) AR-guided capture, a person with a phone follows a trail of chevrons when far from a keyframe and matches its pose when near. In (c) Review and refine, a person holds a tablet that shows the previs and a slider. Two cards below the tablet compare the captured handheld path with the planned target path and show a slider that interpolates between them. A blue arrow labeled revise the same path and timing leads from review back to planning, and a dashed teal loop labeled guided additional capture leads from review back to capture. The right part, titled Previsualize hallucination risk before generating, enlarges the tablet screen. Three previs images sit at slider positions 0.00 (captured, lower risk), 0.67 (intermediate, balances capture and generation), and 1.00 (target, higher risk). Red hatching marks missing coverage and a blue-to-yellow heatmap marks resolution and performance deficits. Both grow from the captured end toward the target end. A card under each image shows the matching camera path and slider position. Four responses sit along the bottom: revise the move, capture more, re-time, and accept hallucination.}
\end{teaserfigure}

\maketitle

\section{Introduction}
\label{sec:intro}

\ad{Previsualization and planning tools have long played a critical role in creative domains that benefit from extensive amounts of attention to detail. Whether it is armature sketching for illustration, storyboarding or animatics for film, or wireframe renders for CGI, the goal is the same,} \nt{to revise decisions before production becomes expensive. In filmmaking, that flexibility traditionally ends when the camera rolls. Executing a complex camera move, such as an orbit, crane, or push-in, requires meticulous choreography, specialized equipment, or skilled operators. Once a take is filmed, altering the camera move traditionally requires another take.}

Generative video reshooting changes this. 
Given a recorded take, these models can render the same performance from a new camera trajectory chosen or revised after capture~\cite{bai2025recammaster,zhang2025recapture,lin2026vista4d}. 
\nt{However, a target trajectory may require rendering things that were not visible at all, or not visible in sufficient detail, in the captured trajectory, leaving parts of the shot for the model to hallucinate. We call this unrecorded demand the shot's \defineterm{\burdenterm{}}. A direct response is to capture additional takes covering alternative camera moves. Yet doing so consumes scarce production time and cannot reproduce the exact moment of a live performance. Moreover, filmmakers may discover that a different move is needed only when editing against adjacent footage. Generative reshooting offers another response: infer the missing content from the recorded take. Research in this area has focused on improving that inference through advances in generative modeling and more extensive use of recorded footage. These improvements, however, do not by themselves help filmmakers decide what additional footage to capture or what content they are willing to let the model hallucinate.} 
\ad{On one extreme we have the challenges of cinematography, and on the other extreme we have the risks of hallucination. 
Navigating that spectrum is going to define a lot of cinematography in the future.}

\nt{In this work, we present \workflowname{}, a workflow that connects physical camera capture and generative reshooting around an editable 3D shot plan. While the plan specifies the camera move the filmmaker wants, a lightweight 3D proxy represents what the take recorded. Evaluating views along a camera move against this proxy produces a previsualization (\previs{}) of \burdenterm{} that updates in real time as filmmakers revise the move. This lets filmmakers rapidly explore moves between the captured and target trajectories, inspecting trade-offs between physical capture and hallucination without waiting for the generative model.}

\nt{We demonstrate this workflow both on set and during editing. On set, a mobile augmented reality (AR) application brings the previsualization into shot planning, guided capture, and review, letting filmmakers revise the move or capture additional footage while the physical scene is still available (Figure~\ref{fig:teaser}). During editing, an offline pipeline applies the same previsualization to existing video, letting filmmakers explore alternative camera moves and decide what hallucination they are willing to accept before committing to generation (Figure~\ref{fig:offline-gallery}).}

\nt{We evaluated \workflowname{} in a two-part study with seven experienced filmmakers across cinematography, directing, and video production. Participants used the previs to revise camera moves and choose targeted capture responses, reporting less pressure to execute a perfect take on set. Their decisions revealed tensions between reducing \burdenterm{} and preserving creative intent, as well as between accepting plausible generated content and maintaining the authenticity of the recorded take.}

Our contributions include:
\begin{itemize}[leftmargin=*,nosep]
    \item \nt{\textbf{\workflowname{}}, a workflow for navigating trade-offs along the continuum between captured and target camera moves, balancing physical capture against generative hallucination (Section~\ref{sec:background}).}
    \item \nt{\textbf{Strategies for visualizing \burdenterm{}} that expose where and why a recorded take lacks support for a target camera move, pairing distinct deficits with targeted responses before generation (Section~\ref{sec:previs}).}
    \item \nt{\textbf{A prototype mobile AR application and an offline pipeline} that demonstrate this workflow across on-set capture and post-production editing (Section~\ref{sec:system}).}
    \item \nt{\textbf{Findings from a study with seven experienced filmmakers} on how they use previsualized \burdenterm{} to make camera decisions and negotiate tensions between hallucination, creative intent, and the authenticity of recorded performance (Section~\ref{sec:evaluation}).}
\end{itemize}

\section{Related Work}
\label{sec:relatedwork}

Prior work in camera planning and video reshooting largely separates capture
from generation. Reshooting models treat the take as fixed and hallucinate whatever a
target move asks for that the take does not contain. Capture guidance and
previsualization, conversely, help filmmakers plan and execute camera moves
without anticipating computational synthesis. Neither tells a filmmaker, while
the scene is still standing, what a move would leave to the model.

\subsection{Generative Video Reshooting}
\label{sec:rw-generative}

Generative reshooting models re-render a recorded take from a new camera
trajectory, conditioning directly on the source video~\cite{bai2025recammaster}
or anchoring generation to geometry rendered from the target
camera~\cite{zhang2025recapture,ren2025gen3c,yu2025trajectorycrafter,paliwal2026reshoot,lin2026vista4d}.
Geometry-based pipelines pass a rendered proxy and a support mask to the
model. This proxy is typically a 3D point cloud or a time-varying 4D
representation, but the mask records only where it has holes. It does not
distinguish a surface that was never observed from one sampled too coarsely for
the requested framing, nor does it describe whether performance evidence is
valid at the requested time. Prior work treats synthesis across those gaps as a
capability: the ReCapture reshooting model~\cite{zhang2025recapture}, for
example, reports that the model can plausibly hallucinate scene content outside
the source view. For a capture workflow, however, that capability is also an
unrecorded production decision: neither the mask nor the generated pixels tell
the filmmaker whether the hallucination was avoidable.

Recent work reduces unsupported generations by routing more of the take
into the result~\cite{paliwal2026reshoot,lin2026vista4d}, completing unseen
geometry with additional models~\cite{cao2026freeorbit4d,jin20264danyonecreate4dcasual},
separating supported from unsupported regions during generation~\cite{hu2026prex},
or supplying appearance from reference images while tying performance to
recorded time~\cite{gwtt}. Completion may improve a render, but for a
performer it is not recovery: a face no camera ever recorded cannot be recovered
by a model that never saw it either. Vista4D~\cite{lin2026vista4d}, which we
adopt as our downstream reshooting model, can incorporate additional scene takes
to reduce hallucination, but it does not tell a filmmaker which observations a
particular reshoot requires. These methods make
increasingly sophisticated use of evidence, but only after capture. None of
them evaluate \burdenterm{} while a filmmaker can still record the missing
support.

\subsection{Shot Authoring and Previsualization}
\label{sec:rw-previz}

Previsualization lets filmmakers evaluate framing, camera movement, and
performance before committing production resources. In virtual cinematography,
early work automated camera placement and pacing by encoding film idioms as
procedural rules~\cite{he1996virtual}, while subsequent tools gave authors
interactive control over viewpoint selection through geometric
constraints~\cite{toricspace,directorlens}. CollageVis brings recorded acting
into this process, arranging segmented performers as video layers on a virtual
stage~\cite{collagevis}. This preserves the performance without requiring
character animation, but showing the acting from another direction requires
another take. Generative reshooting offers a way to view a recorded performance
from new directions, but raises the opposite problem: asking for views the take
never recorded.

PrevizWhiz uses rough 3D scenes and video references to guide generative
video previsualization~\cite{previzwhiz}, providing resemblance controls to
steer adherence to the authored scene and motion controls to adjust performance
detail. But PrevizWhiz begins from virtual designs and text prompts, where
hallucinated details are part of ideation. In a live-action take, hallucination
can compromise the authenticity of a recorded performance. Other systems
connect shot authoring directly to physical capture and subsequent editing
through shared representations. CineCraft uses a persistent storyboard-derived
shot plan to coordinate AR guidance, focus and zoom automation, and rough-cut
assembly~\cite{cinecraft}. Hybrid Tours frames a spectrum between all-real and
all-virtual workflows, letting users pick a point on it by filtering and
combining short captured clips into longer camera
trajectories~\cite{hybridtours}. We take that spectrum as an interaction target.
Where Hybrid Tours has users choose among clips they captured, we make the
range between the \capturedmove{} and the \targetmove{} continuously navigable
within a single take, attaching a real-time readout of the \burdenterm{} each stop
incurs. More broadly, while these systems carry one representation across
production stages, the take is always what a later stage consumes. It is never
what the plan is measured against. Our \previs{} makes that measurement
visible as filmmakers change the camera path and its timing.

\subsection{Capture Guidance}
\label{sec:rw-guidance}

View synthesis and reconstruction methods prescribe where to capture and
how densely to sample a
scene~\cite{davis2012unstructured,locallightfield,yasunaga2025intellicap}. For
camera operation, ARCAM visualizes camera movement in the physical scene so
novices can reproduce a move demonstrated in a sample video~\cite{arcam}, and
LookOut combines planned framing with responsive gimbal control during a
take~\cite{lookout}. A second group guides a user back to a view that was
already recorded: the ReCapture AR time-lapse system~\cite{yan2024recapture}
returns users to earlier viewpoints so hand-held time-lapse can be shot one frame
at a time, and Personal Time-Lapse adds interactive feedback for re-establishing
a viewpoint, body pose, and lighting condition~\cite{mecapture}.

In all of this work the view to acquire is known before guidance begins,
derived from a sampling rule, a reference video, or a frame the user already
shot. Our workflow has to derive it first: the \targetmove{} and the support it
lacks determine which view or performance is needed, and only then does guidance
have somewhere to point. This derivation also turns execution into a choice,
because the filmmaker can revise the move, capture more, or accept
hallucination. Camera departures from the plan during capture are not errors.
Instead, they provide opportunistic coverage that can support alternative
moves. When capture is chosen, the new recording is checked against the same
\targetmove{}, keeping both the move and its supporting evidence open to
revision.

\subsection{Steering and Inspecting Generative Assistance}
\label{sec:rw-humanai}

Mixed-initiative interfaces allow computational systems to assist without
usurping user control, proposing rather than acting and keeping the basis for
each proposal open to inspection~\cite{horvitz1999principles}. Human--AI
interaction guidelines similarly emphasize that a system should make clear what
it can do and how well it can do it, ensuring reliance tracks competence rather
than aesthetic appearance~\cite{amershi2019guidelines}. Generative reshooting
makes this unusually difficult: the output is a finished-looking video, and
nothing in it separates the parts that reproduce the take from the parts the
model hallucinated.

Creative tools have responded with controls over where delegation happens.
In virtual cinematography, XCam lets operators adjust automation separately for
object tracking, camera motion, and scene transitions~\cite{xcam}. Others
provide structures for inspecting and revising AI assistance, either through
interconnected representations across narrative planning and timeline
editing~\cite{videorigami}, or by aligning alternative AI-assisted edits so
differences can be compared without reviewing unchanged
content~\cite{videodiff}. Our workflow similarly leaves decisions about what to
generate to the filmmaker. But whereas these systems evaluate or constrain the
model's output after generation, our \previs{} acts on its input: filmmakers
adjust what the model will receive, seeing the consequence on an interactive
\proxy{} before a single frame is generated.

\section{\Workflowname{} Workflow}
\label{sec:background}

We call \workflowname{} a workflow rather than an interactive system because it changes the process of reshooting, not one tool within it. Today a filmmaker records a take, leaves the set, chooses a \targetmove{}, generates, and inspects the result for what the model had to hallucinate. \workflowname{} checks the take's support for a move before generation, from a \proxy{} of the take, which makes the check cheap enough to repeat and early enough to capture more:

\begin{table*}[t]
\centering
\small
\begin{tabular}{@{}llll@{}}
\toprule
\textbf{Workflow} & \textbf{Missing evidence found} & \textbf{Cost of a check} & \textbf{Responses} \\
\midrule
Generate and inspect & After generation & One generation & Revise the move, regenerate \\
\workflowname{}, on set & Before generation & A \proxy{} render & Revise, capture more, accept \\
\workflowname{}, in editing & Before generation & A \proxy{} render & Revise, accept \\
\bottomrule
\end{tabular}
\end{table*}

Filmmakers already have conventions for deciding what to shoot and what to leave to post. Camera overlays such as zebra stripes and focus peaking report what the sensor recorded and leave the judgment to the operator. Editors cut on low-resolution proxies because the full render is the expensive step. Crews shoot pickups to cover what a take missed, block to a shot list and marks on the floor, and record what will be fixed in post.

\begin{figure*}
  \centering
  \includegraphics[width=\linewidth]{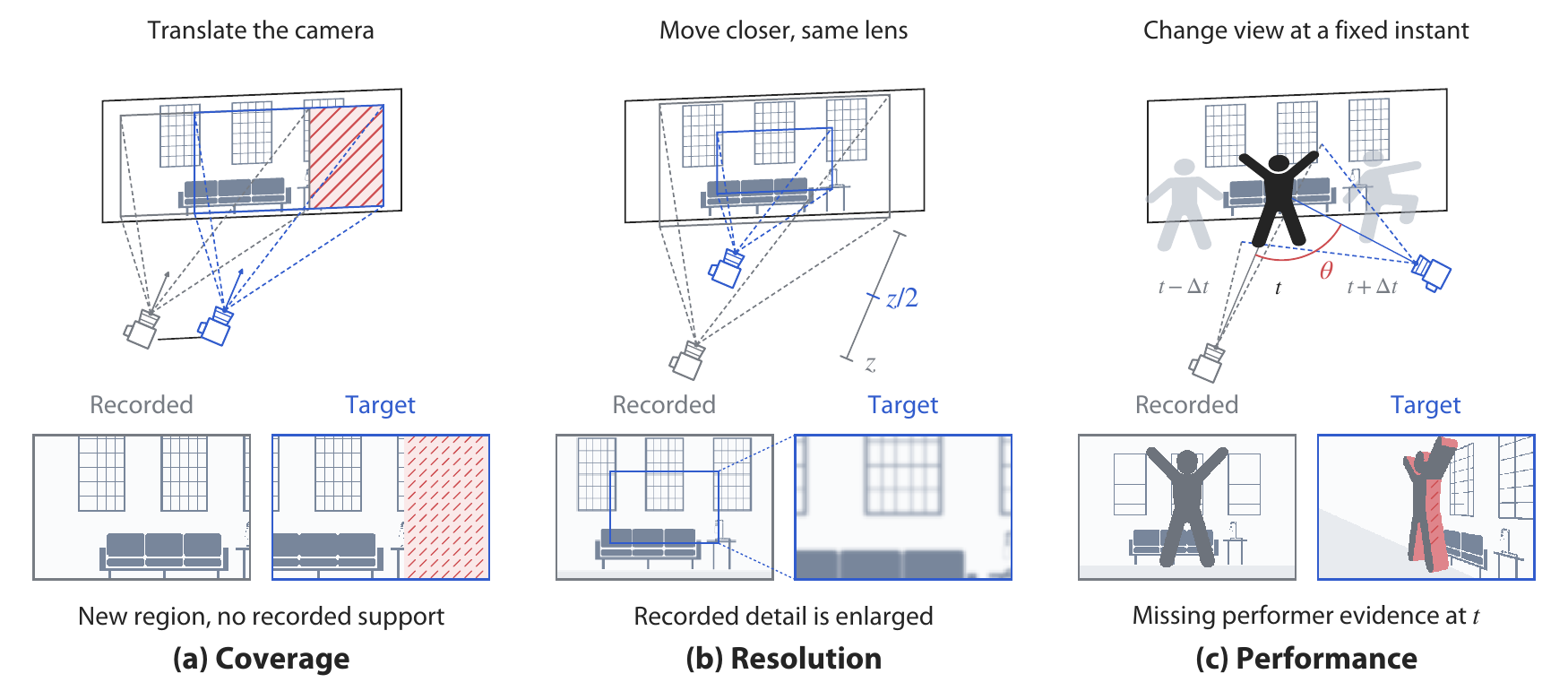}
  \caption{Our \burdenterm{} previs estimates three deficits that vary
  with edits to the target camera move and indicate what downstream reshooting
  models must synthesize: (a) \textbf{coverage} for surfaces unseen from a compatible
  direction, (b) \textbf{resolution} for insufficient detail for the target
  framing, and (c) \textbf{performance} for missing performer evidence from a
  compatible viewpoint at the target instant.}
  \label{fig:signal_geometry}
  \Description{Three side-by-side constructions with recorded and target
  image pairs underneath. The scene contains three windows, a sofa, and a lamp. In coverage,
  parallel gray and blue cameras face the wall. The target
  translates to the right without rotating, and a newly revealed wall region is hatched
  red in both the construction and target image. In resolution, a target
  camera moves halfway toward the wall. A compact, unconnected distance
  scale in the white space beside the frustums marks the target depth z/2
  halfway to the recorded depth z. A blue rectangle in the recorded
  image connects to the enlarged target crop, whose recorded samples are
  visibly softer while the recorded image remains sharp. In performance,
  a performer stands in front of the same
  wall with arms raised. A faded figure with lowered arms precedes it and
  a faded figure with a lifted knee follows it. Each time label sits below
  its pose, clear of the background. Gray and blue rays connect the
  selected instant to two cameras, whose images both show the raised-arm
  pose. The target image marks performer surfaces
  absent from the recorded view at that instant.}
\end{figure*}

Our design goals ground \workflowname{} in these conventions, in the capture deficits defined in Section~\ref{sec:previs} (coverage, resolution, and performance, Figure~\ref{fig:signal_geometry}), and in the cost of generating a shot, which is too slow to iterate with on set or in editing:

\begin{itemize}[leftmargin=*,nosep]
  \item \textbf{DG1: Previsualize \burdenterm{} before generating.} Before
  anything is generated, the \previs{} should show where in the frame and
  when in the shot the take fails to support a move, and why, computed
  from the take rather than from the model's output.

  \item \textbf{DG2: Make the range between the captured and target moves
  navigable.} Between the \capturedmove{} and the \targetmove{} lies a
  range of possible moves, each asking the model to hallucinate more or
  less. The filmmaker should be able to navigate quickly through that
  range and see the \burdenterm{} for every move in it, on set or in
  editing, without calling the model.

  \item \textbf{DG3: Map each deficit to its own previs signal and its own
  response.} Filmmakers should be able to tell each kind of deficit apart
  and pair it with the response that fixes it.

  \item \textbf{DG4: Keep one editable shot plan throughout.} The shot
  plan, the \targetmove{} as keyframes and timing, should be the one
  object every step works from. It guides the take, it is what the take
  is measured against in review, it routes a pickup, and it is what the
  model receives. It should stay editable at each of these points, with
  every change updating the \previs{}.

  \item \textbf{DG5: Let the filmmaker decide what to leave to the model.}
  Whether a region may be left to the model depends on what it means to
  the shot, which the geometry cannot tell. The filmmaker should make that
  call knowingly.
\end{itemize}

Section~\ref{sec:previs} formalizes these capture deficits and their computation on a 3D proxy, and Section~\ref{sec:system} demonstrates the resulting workflow in a mobile application (Figure~\ref{fig:teaser}) and on ordinary video.

\section{Previsualizing \BurdenTerm{}}
\label{sec:previs}
\label{sec:system:evidence}
\label{sec:system:conditioning}
\label{sec:system:overlays}

Evaluating \burdenterm{} requires measuring what a move asks for that the
recorded take cannot support. Our \previs{} checks this support directly against
a lightweight 3D \proxy{} of the take in milliseconds, before running any
generative model (DG1). In this section, we define the three distinct deficits
that arise between a move and a take, formalize the continuous range of moves
between the \capturedmove{} and the \targetmove{}, and describe how these
deficits are computed and visualized.

\subsection{What a Take Must Supply for a Move}
\label{sec:background:burden}

\Burdenterm{} depends on both the viewpoint a move requests and the visual
evidence a take provides. At output time $t$, the target camera $\camtar(t)$ asks
for a view of the scene and performance at source time $t_s$. We decouple
camera timing from performance timing while keeping the source clock unchanged
($t_s = t$): editing camera pacing changes when the camera reaches each
viewpoint, while the recorded performance proceeds at its original speed.

In our experiments with recent generative reshooting models, holding a take fixed
and varying viewpoint, framing, or timing reveals three recurring capture deficits
(Figure~\ref{fig:signal_geometry}). First, a surface may never have been
observed from a compatible viewing direction (\emph{coverage}). Second, even
when observed, it may be sampled too coarsely for the requested framing
(\emph{resolution}). Third, while static scene evidence accumulates across the
entire take, a moving performer's appearance is tied to a specific instant. A
different frame of an actor does not record the same physical action
(\emph{performance}).

\subsection{From Deficits to a Previs}
\label{sec:previs:previs}

To evaluate these deficits without running the generative model, we construct a 3D \proxy{} of the take consisting of an accumulated static point cloud and time-indexed performer point clouds (DG1). 
Rendering this \proxy{} along a move produces target-view point projections and binary support masks that match the conditioning representations required by downstream reshooting models such as Vista4D~\cite{lin2026vista4d}. The \previs{} then tests the observations backing each rendered point against our three deficits in milliseconds, enabling instant feedback as filmmakers edit a move.

To make the continuum between the \capturedmove{} and \targetmove{} navigable (DG2), we interpolate between the captured camera $\camcap(t)$ and target camera $\camtar(t)$ at each output time $t$:
\begin{equation}
\camlam(t)=\operatorname{mix}\bigl(\camcap(t),\camtar(t),\lam\bigr),\qquad \lam\in[0,1],
\label{eq:family}
\end{equation}
where $\operatorname{mix}$ interpolates 6-DoF camera pose and focal length. At
$\lam=0$, the camera traces the \capturedmove{}, where deficits reflect only
sensor occlusions or dropouts, and at $\lam=1$, it reaches the full
\targetmove{}.

For each projected point rendered from $\camlam(t)$, the \previs{} evaluates support:
\begin{itemize}[leftmargin=*,nosep]
  \item \textbf{Coverage} holds when at least one source observation falls within an angular cone around the target viewing direction.
  \item \textbf{Resolution} evaluates the ratio between the requested sampling density (which increases as the camera approaches or the lens tightens) and the densest compatible source observation. A ratio exceeding one requires the model to hallucinate detail absent from the take.
  \item \textbf{Performance} is evaluated per instant rather than per pixel. It measures the fraction of visible performer samples in the target view supported by source observations at instant $t$ from a compatible angle.
\end{itemize}
Where tracking drops out or depth is missing, the \previs{} explicitly reports \emph{unknown} rather than false support.

\begin{figure}
  \centering
  \includegraphics[width=\linewidth,height=0.8\textheight,keepaspectratio]{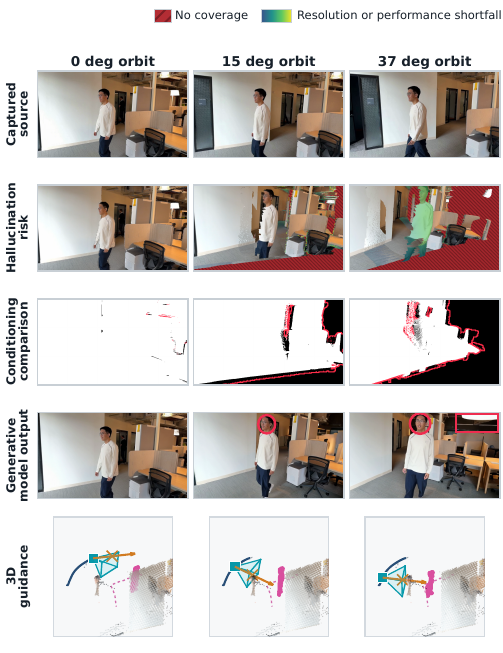}
    \caption{\nt{\textbf{Comparing the \previs{}, model conditioning, and generated output.} Columns show the same take reshot as 0$^\circ$, 15$^\circ$, and 37$^\circ$ orbits. From top to bottom: the source frame, the \previs{} showing \burdenterm{} (hatched red where no frame of the take saw the pixel, Viridis tint where resolution or performance falls short, brighter as the shortfall grows), Vista4D's conditioning mask (white is conditioned) with our coverage estimate outlined in red, the generated output with synthesized regions circled, and pickup guidance. Wider orbits expose more missing content. The \previs{} locates missing support before generation. It does not predict the quality of the synthesized content.}}
  \label{fig:signal_correspondence}
  \Description{Five rows by three columns. The columns are 0, 15, and 37 degree orbits of the same take. The rows show the captured frame, the previs with hatched red for missing coverage and a Viridis tint on the performer and room edges that grows with the orbit, the model's binary conditioning mask with a red outline where our coverage estimate overlaps its unconditioned pixels, the generated frames with circled synthesized regions, and a 3D view of the room cloud with the pickup camera frustum and aim ray.}
\end{figure}

The resulting signals appear as overlays in the target view and as summary lanes along the timeline (DG3, Figure~\ref{fig:review_interface}). In addition, a 3D map colors the interpolated trajectory manifold by \burdenterm{}, letting filmmakers compare alternatives across $\lam$ with a slider. Figure~\ref{fig:signal_correspondence} compares these diagnostic signals with
Vista4D's conditioning masks and generated outputs across three orbit angles.
The signals locate missing support rather than predict how convincingly the
model will hallucinate.

Keeping these signals distinct is critical because each maps to a different production decision (DG3): capturing more through a later pickup can supply static coverage or resolution, but cannot reconstruct an unrecorded moment of live performance. Section~\ref{sec:system:respond} translates these diagnostic distinctions into actionable controls on set (DG5).

\section{Workflow Implementation: App and Offline Pipeline}
\label{sec:system}

We demonstrate \workflowname{} across two settings: a mobile AR application for
on-set production and an offline pipeline for ordinary video
(Figure~\ref{fig:teaser}). Following the principle that one shared
representation should coordinate production stages~\cite{cinecraft,hybridtours},
the workflow anchors each stage in the editable 3D shot plan checked against a 3D
\proxy{} of the take (DG4). We describe the on-set workflow below, and detail the
end-to-end data pipeline in Appendix~\ref{app:dataflow}
(Figure~\ref{fig:dataflow}).

\subsection{Planning the Move in the Scene}
\label{sec:system:plan}
\label{sec:system:pathtiming}

\begin{figure*}[p]
  \centering
  \includegraphics[width=\linewidth]{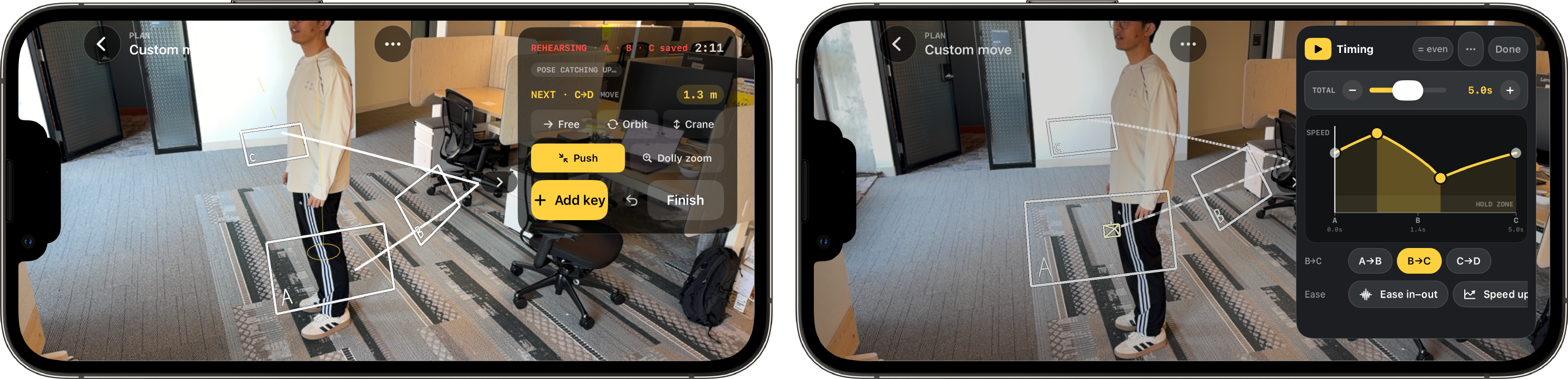}
  \caption{\textbf{Authoring a move in AR.}
  \textbf{Left}: path. The filmmaker walks to each viewpoint and drops an
  AR keyframe. Each stretch between keyframes is a free move by default.
  Tapping a common shot-type preset, such as an orbit, crane, push, or dolly zoom, instead generates the
  keyframes for that move, and the system joins consecutive stretches into
  one continuous path.
  \textbf{Right}: timing. A speed curve over the same path defines camera
  velocity and holds. A double tap adds a control point, dragging it up or down
  speeds up or slows that stretch, and pulling it to the axis holds the camera
  still. Presets such as ease in-out apply a common profile to a stretch.}
  \label{fig:authoring}
  \Description{Two landscape phone screenshots. Left: an AR view of a
  performer in an office with connected camera keyframes A, B, and C,
  and a panel offering Free, Orbit, Crane, Push, and Dolly zoom, with Push
  selected. Right: the same scene with a Timing panel showing a speed
  curve over keyframes A through C, a hold zone along the horizontal axis,
  a total duration of five seconds, and per-stretch easing options.}
\end{figure*}

To keep the shot plan editable throughout production (DG4), the filmmaker authors it directly in the physical scene. Walking through the environment, the filmmaker frames key viewpoints and
saves AR camera poses (Figure~\ref{fig:teaser}a and Figure~\ref{fig:authoring},
left). The app interpolates a smooth camera path with $C^1$ continuity in position and velocity (Appendix~\ref{app:splines}) through freeform keyframes or shot presets (orbit, crane, push, dolly-zoom), previewing planned views with point-based rendering before rolling.

Timing is controlled independently (Figure~\ref{fig:authoring}, right). A
timing curve defines camera velocity and holds along the path, initialized from
the pacing of the filmmaker's rehearsal walk. Decoupling path from timing allows
retiming the move against the recorded performance without re-walking the shot
(Section~\ref{sec:background:burden}). During post-capture review, keyframes can
be adjusted in the interface or re-anchored by standing in the scene and saving a
new AR pose.

\subsection{Capturing With the Plan as Guidance}
\label{sec:system:capture}

\begin{figure*}[p]
  \centering
  \includegraphics[width=\linewidth]{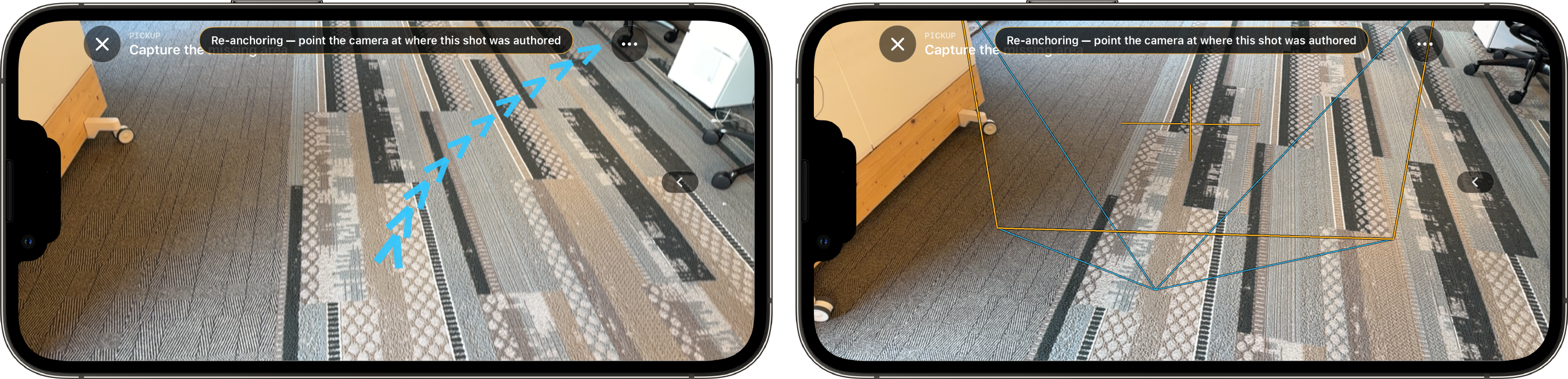}
  \caption{\textbf{Coarse-to-fine AR guidance for capture, shown during a pickup.}
  \textbf{Left}: coarse guidance. A trail of arrows on the floor leads the
  operator toward the saved viewpoint.
  \textbf{Right}: fine guidance. The saved camera's frustum and a central
  crosshair help the operator match its position and framing.
  The same guidance supports rehearsal, the take, and pickup views saved
  during review.}
  \label{fig:guidance_pickup}
  \Description{Two landscape phone screenshots of the Pickup interface in
  an office. Left: blue arrows form a trail across the floor toward a saved
  viewpoint. Right: a blue camera frustum, a yellow rectangular outline,
  and a yellow crosshair overlay the scene. Both show a banner asking the
  operator to point the camera at where the shot was authored to re-anchor
  the plan.}
\end{figure*}

During a take, the shot plan acts as virtual marks on the floor, guiding camera execution in 3D space (DG4, Figure~\ref{fig:teaser}b). To prevent visual clutter during capture, the interface employs \emph{progressive disclosure} (Figure~\ref{fig:guidance_pickup}). When the operator is far from the next keyframe, \emph{coarse guidance} projects breadcrumb arrows on the ground indicating where to move. As the operator approaches, these arrows give way to \emph{fine guidance}: the target camera's frustum, distance readouts, and framing cues to settle the shot. Guidance applies equally to rehearsal walks and pickup tasks (Section~\ref{sec:system:respond}).

Importantly, guidance does not enforce rigid adherence. If the operator
departs from the planned path, the take is not discarded. Instead, review
evaluates whatever evidence the camera recorded. This allows operators to focus
on capturing the scene rather than chasing mechanical precision. During capture, the device records RGB video, hardware LiDAR depth, and ARKit 6-DoF camera poses, while tracking performers in real time using EdgeTAM~\cite{zhou2025edgetam}. These streams populate the static and dynamic stores of the \proxy{} for immediate on-device review.

\subsection{Reviewing and Responding on Set}
\label{sec:system:review}
\label{sec:system:respond}

\begin{figure*}[t]
  \centering
  \includegraphics[width=\textwidth]{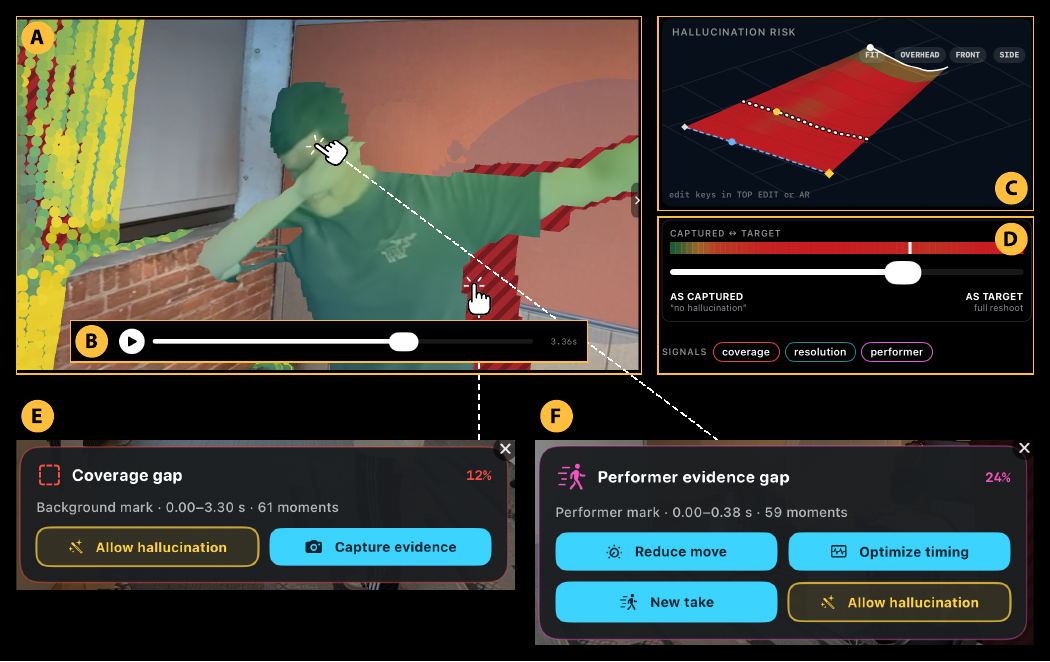}
  \caption{\textbf{Review and Refine view.}
  (A) The \previs{} overlays the 3D proxy with \burdenterm{}:
  red hatching marks missing coverage, while a heatmap shows resolution and performance
  deficits. (B) controls playback. Toggling signals updates the overall risk map.
  (C) shows the captured and target trajectories and the interpolated moves
  selected with the compromise slider (D). Tapping a flagged region opens a
  popup with remedy strategies for (E) coverage or resolution gaps and
  (F) performance evidence gaps.}
  \label{fig:review_interface}
  \Description{Six labeled panels from the Review interface. The upper row has a large
  previs picture on the left with red hatched regions at the right edge and
  a transport bar labeled B overlaid near its bottom edge. On the right are
  a 3D risk map labeled C and a slider, risk strip, and signal controls labeled D.
  Below, E is a Coverage gap response card with Allow hallucination and
  Capture evidence buttons. F is a Performer evidence gap response card with
  Reduce move, Optimize timing, New take, and Allow hallucination buttons.
  Dashed white lines connect E to a red-hatched coverage region and F
  to the performer, meeting the bottom of each tap icon’s palm. Opaque white tap icons mark both selected regions.
  The coverage connector passes behind the timeline.
  Gold circles label the panels.
  The confirmation step is not shown.}
\end{figure*}

Review brings the target view, timeline, compromise slider, and 3D risk
map together on a single screen (DG1, DG2, Figure~\ref{fig:review_interface}).
Tapping a flagged region opens an actionable response card tailored to that
deficit (DG3, Figure~\ref{fig:review_interface}E--F). Rather than prescribing an
automated fix, the interface supports three explicit production actions (DG5):

\begin{itemize}[leftmargin=*,nosep]
  \item \textbf{Revise the move.} The filmmaker can edit keyframes, adjust the
  timing curve, or inspect two automated proposals. \emph{Reduce move} pulls the slider back toward the \capturedmove{} until the flagged region is supported. \emph{Retime} searches for a camera velocity schedule that slows down where the take provides support and accelerates through unsupported spans, while preserving authored holds (Appendix~\ref{app:retime}). Proposals are visualized with their trade-offs and never applied automatically (DG5).
  \item \textbf{Capture more.} For coverage or resolution deficits, tapping the card creates a targeted pickup task. The app freezes the unsupported region and time span and extracts one or two guidance viewpoints. The operator then captures a brief pickup take, which is folded into the \proxy{} and immediately re-evaluated to verify that the gap has been closed.
  \item \textbf{Accept hallucination.} When a missing region is deemed artistically inconsequential or easily synthesized (such as an unfeatured background wall), the filmmaker explicitly confirms it (DG5). The interface records an acceptance mask tied to the current shot plan. Any subsequent edit to the move or take clears the acceptance, ensuring that only deliberate decisions reach post-production.
\end{itemize}

What leaves the set is the finalized camera move, verified pickup takes, and confirmed acceptance masks for the generative reshooting model.

\subsection{The Previs on Ordinary Video}
\label{sec:system:offline}

The \previs{} does not require custom AR hardware during capture. An offline pipeline ingests ordinary video, estimates camera poses and dense depth using $\pi^3$~\cite{wang2026pi} or Depth Anything~3~\cite{lin2025depth}, and extracts performer masks using SAM~3~\cite{carion2026sam}. From these inputs, it constructs the 3D \proxy{} and serves it to a browser-based viewer with the same slider, timeline lanes, and diagnostic overlays (DG2). An interactive build of this viewer is included in the supplemental material, allowing users to scrub the compromise slider and inspect diagnostic overlays across sample sequences.

\begin{figure*}[p]
\centering
\includegraphics[width=\linewidth]{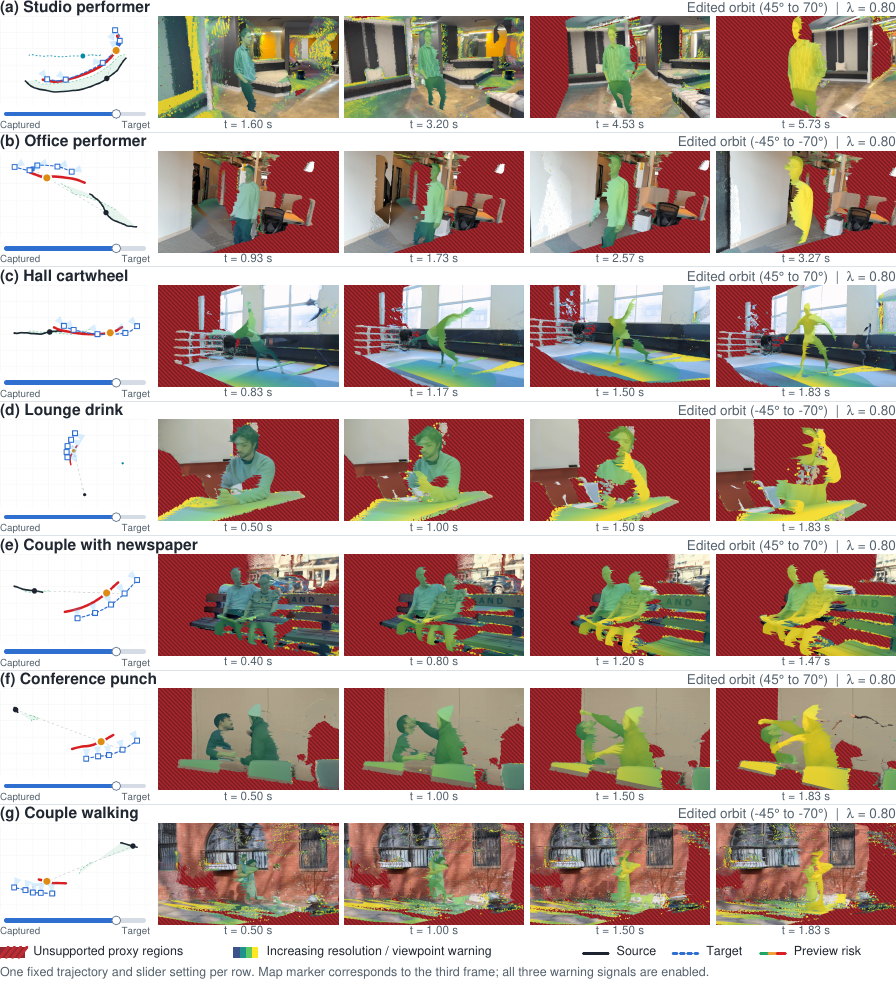}
\caption{\textbf{The \previs{} on ordinary video.} Seven scenes recorded as RGB video without our application and reconstructed offline, with camera poses and depth from monocular estimation (Pi3X) and performer masks from video segmentation (SAM3), then reviewed in a browser viewer with the same map, keyframes, and slider as the app. Each row shows the captured path in black, an edited orbit as the target in dashed blue, the preview risk of the selected move at $\lam=0.8$, and four successive frames of the \previs{} for that move. Every orbit swings toward a side of the performer the take never filmed, so the performer wash rises toward yellow as the arc passes that side, while red hatching marks background the take never recorded and the same ramp on the background marks detail it sampled too coarsely. These are renders of what the take supports, not generated reshoots.}
\label{fig:offline-gallery}
\Description{Seven horizontal comic strips, one per scene: a studio performer, an office performer, hall cartwheel, lounge drink, a couple with a newspaper, conference punch, and a walking couple. Each strip starts with the actual camera-trajectory map and a captured-to-target slider, followed by four chronological reconstructed frames with warning overlays and timestamps. The map marker corresponds to the third frame. Red hatching marks unsupported proxy regions; blue-green through yellow tints indicate increasing resolution or performer-viewpoint warnings. Each row uses an edited target trajectory that moves around the performer to reveal unseen sides, at a fixed slider setting.}
\end{figure*}

Figure~\ref{fig:offline-gallery} demonstrates seven ordinary video sequences with edited camera moves revealing unseen angles of performers. When the original scene and actors are no longer available for physical pickups, filmmakers rely on the \previs{} to balance creative intent against \burdenterm{}, choosing whether to revise the move or accept hallucination before running the generative model.

\section{Evaluation}
\label{sec:evaluation}

\begin{figure*}[t]
\centering
\includegraphics[width=\textwidth]{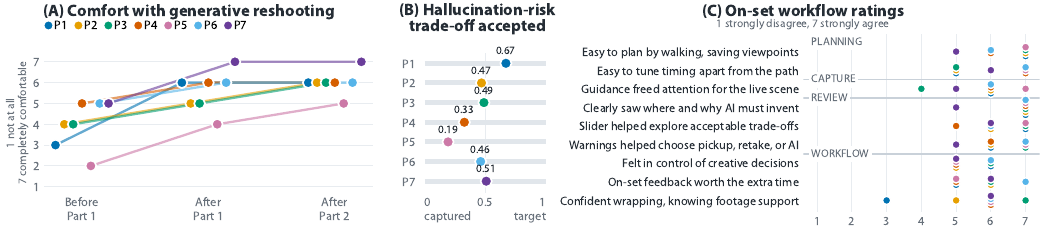}
\caption{Study results, one color per participant.
(A) Comfort with generative video reshooting before Part~1, after Part~1,
and after Part~2.
(B) Where each participant stopped on the Part~1 slider, between the
\capturedmove{} at 0 and the \targetmove{} at 1. (C) Ratings of the on-set workflow after Part~2, one
row per item. Item wording is abbreviated. The
questions as asked are in Appendix~\ref{app:study}.}
\label{fig:study-comfort}

\Description{Three panels. Panel A shows comfort at three points per
participant. P1 3, 6, 6. P2 and P3 4, 5, 6. P4 5, 6, 6. P5 2, 4, 5. P6 5, 6, 6. P7 5, 7, 7.
Medians 4, 6, 6. Panel B shows one row per participant on a track from 0,
the captured move, to 1, the target move, with accepted stops at 0.67, 0.47,
0.49, 0.33, 0.19, 0.46, and 0.51 in participant order. Panel C shows nine rating
items with one marker per participant on a 1 to 7 scale, in participant order.
Easy to plan by walking and saving viewpoints 7, 7, 7, 6, 7, 6, 5. Easy to tune
timing apart from the path 5, 5, 5, 7, 7, 7, 6. Guidance freed attention for the
live scene 6, 6, 4, 6, 7, 6, 5. Clearly saw where and why AI must invent 7, 7, 7,
7, 7, 7, 5. Slider helped explore acceptable trade-offs 7, 6, 7, 5, 7, 6, 6. Warnings
helped choose pickup, retake, or AI 6, 6, 7, 6, 7, 7, 5. Felt in control of
creative decisions 6, 5, 6, 5, 5, 6, 5. On-set feedback worth the extra time 5, 6,
6, 5, 5, 7, 6. Confident wrapping, knowing footage support 3, 5, 7, 6, 6, 6, 6.}
\end{figure*}

We evaluated the \workflowname{} workflow with seven experienced filmmakers to examine how practitioners read \burdenterm{} and navigate production trade-offs. Because this approach introduces an unprecedented coupling between physical camera capture and generative reshooting, no direct baseline exists. Following prior evaluations of interactive creative tools without direct counterparts~\cite{previzwhiz,cinecraft,collagevis,realitysketch,pronto}, we report qualitative sessions examining how filmmakers reason about shot planning, guided capture, and \burdenterm{} across production and post-production.

\begin{table*}[p]
\centering
\caption{Film and video production backgrounds of the participants.}
\label{tab:rewrite-participants}
\small
\renewcommand{\arraystretch}{1.1}
\begin{tabular}{@{}p{0.04\textwidth}p{0.06\textwidth}p{0.10\textwidth}p{0.24\textwidth}p{0.20\textwidth}p{0.23\textwidth}@{}}
\toprule
\textbf{ID} & \textbf{Gender} & \textbf{Experience} & \textbf{Primary Roles} & \textbf{Production Domain} & \textbf{VFX and Generative AI} \\
\midrule
P1 & M & 9 years &
\raggedright Photographer, videographer &
\raggedright Social and brand content &
\raggedright Prior VFX collab with Disney, Pixar, ILM/Lucasfilm. Gen AI: Veo, Nano Banana \tabularnewline
P2 & M & 6.5 years &
\raggedright Cinematographer (DP), grip, editor &
\raggedright Student and indie films &
\raggedright Rare VFX. Gen AI: open-source generative video models \tabularnewline
P3 & F & 4 years &
\raggedright Video producer &
\raggedright Commercial and advertising &
\raggedright Occasional VFX, 3D modeling. Gen AI: Midjourney \tabularnewline
P4 & M & 8 years &
\raggedright Cinematographer (DP), AC, gaffer, grip &
\raggedright Narrative short films &
\raggedright Never VFX. Gen AI: Midjourney, Topaz \tabularnewline
P5 & F & 13 years &
\raggedright Cinematographer (DP), Assistant Camera (AC) &
\raggedright Narrative and indie films &
\raggedright No VFX. Gen AI: ChatGPT, Adobe AI tools \tabularnewline
P6 & M & 7 years &
\raggedright Cinematographer (DP), Assistant Camera (AC) &
\raggedright Film, commercial shorts, documentaries, dance videos &
\raggedright VFX for extensions and cleanup. Gen AI: Seedance \tabularnewline
P7 & M & 15 years &
\raggedright Director, teaching professor &
\raggedright Commercials, music videos, installations &
\raggedright VFX in post with editors and VFX teams. Gen AI: Veo, Runway, Nano Banana, ComfyUI, Google Flow \tabularnewline
\bottomrule
\end{tabular}
\end{table*}

\subsection{Participants}

We recruited seven experienced filmmakers (5 male, 2 female,
Table~\ref{tab:rewrite-participants}) through professional networks and filmmaking communities, with four to fifteen years of production experience. Three work as directors of photography (DPs, or cinematographers) on narrative and independent films, one produces advertising video, one shoots brand content, one directs and shoots documentaries and shorts, and one directs commercial campaigns and music videos. Prior use of generative models ranged from none to regular production use. Sessions lasted 45 to 60 minutes, and participants were compensated for their time.

\subsection{Study Design}

\begin{figure*}[t]
\centering
\includegraphics[width=\textwidth]{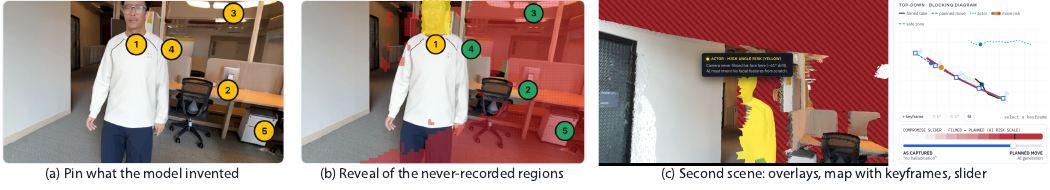}
\caption{Part~1 interactive browser exercise. (a) Participants pin what they think a reshooting model invented in a generated view. (b) The reveal marks what the camera never recorded. (c) Participants navigate the range of possible reshoots with the \previs{}, the map, and the slider. The green shaded region in the map marks the range of camera trajectories estimated to remain below the shot-support risk threshold.}
\label{fig:study-part1}
\Description{Three screenshots of the browser study. Left, a generated front view of a performer in an office with five numbered pins placed by a participant. Middle, the same view with a red hatch over never-recorded surfaces and a yellow mask over the face. Right, the study scene two thirds of the way through the take with the slider two thirds toward the planned move: the reconstructed take with the coverage hatch and the performer washed yellow, with a popup reading that the camera never filmed his face here and the AI must invent his facial features, beside a top-down map with keyframes and the compromise slider.}
\end{figure*}

Part~1 introduces participants to \burdenterm{} in a browser viewer using an offline take before working on set. Participants first rate how comfortable they would be using a generative model to add a new camera move to their own recorded takes, providing the baseline for Figure~\ref{fig:study-comfort}A. A tutorial sequence introduces the interactive slider and \previs{}. An initial inspection task pairs a take filmed from the side with a front view synthesized by a reshooting model, asking participants to place pins where they believe the model had to invent content (Figure~\ref{fig:study-part1}a). A reveal then overlays the diagnostic map of what the camera never recorded, confirming their intuition by highlighting unobserved background surfaces in red and the performer's face in yellow (Figure~\ref{fig:study-part1}b). The second scene moves participants into the \previs{} reconstructed from the take, using an orbit around the performer as the \targetmove{} (Figure~\ref{fig:study-part1}c). The slider lets them move between the \capturedmove{} and \targetmove{} to inspect support. After modifying the \targetmove{} and exploring the \previs{}, participants stop at an acceptable slider position and explain the trade-off. Finally, they rate their comfort a second time.

Part~2 moves participants onto the set, where our mobile application demonstrates the full workflow combining shot planning, guided capture, and review against the \previs{}. Following a fifteen-minute tutorial, participants plan concatenated paths in augmented reality and customize timing along each. They capture three scenes, including one with a performer and one with a prop. For each take, the experimenter acts as director and requests a change after capture, such as speeding through one stretch and slowing through another. Participants refine the shot in the app, editing path and timing, and navigating between revising the move, retiming, capturing more, or accepting synthesis via the slider while thinking aloud throughout. Finally, they complete a questionnaire on planning, capture, review, and the workflow as a whole (Figure~\ref{fig:study-comfort}C).

\subsection{Findings}

We organize our findings around four themes that reflect the production workflow: (1) authoring and guided capture with a unified shot plan, (2) diagnosing \burdenterm{} and navigating trade-offs before generation, (3) selecting targeted responses for distinct deficits, and (4) how filmmaker comfort grew with active agency and control.

\subsubsection{Carrying a Unified Shot Plan from Preproduction Through Capture} 
Planning a camera path with AR keyframes was rated easy, with a median of 7 of 7 (Figure~\ref{fig:study-comfort}C). P4 observed that walking keyframes felt ``directly transferable'' to how a DP plans a scene by placing the camera where a move starts and ends. On set, guidance replaced marks on the floor (P5), was more precise than a verbal walkthrough (P1), and made physically difficult moves feasible (P2). 

Different production backgrounds shaped distinct mental models for camera planning. While cinematographers like P4 anchored planning to physical marks on set, P3 (who brought a 3D animation background, rating guidance 4 of 7) expected keyframes to follow 3D timeline conventions, noting that guidance meant she would ``only need to capture key frames instead of get the precise movement.'' Beyond individual operation, unifying authoring and capture into an editable plan (DG4) provided a shared reference for crew communication (P1, P2, P6). P1 described the AR path as ``almost like a 3D storyboard'' that makes it ``much easier to convey the vision and timing for a shot across a small-production crew,'' replacing verbal walkthroughs (P6) and helping crews get on board with complex moves (P2).

Having pickups available changed how participants shot takes. Relieved of single-take pressure, operators felt freer moving the camera (P2) and, as P5 put it, ``more free to make mistakes,'' because the workflow ``allows for preplanning to occur without too much pressure for the actual moment.'' Operators prioritized live performance over background coverage (P2), used the visual path to rehearse marks (P3), and noted its value for unrepeatable documentary moments (P5).

\subsubsection{Reading Hallucination Risk and Navigating Trade-offs Before Generation} 
Previsualizing \burdenterm{} before generation (DG1) made model dependencies immediately interpretable. All participants correctly identified the hallucinated face in the initial inspection task before overlays were explained, and rated the \previs{} a median of 7 of 7 for showing where and why the model must invent content (Figure~\ref{fig:study-comfort}C). Participants valued knowing on set what the footage supported (P2, P3, P4, P5, P6, P7). P2 was ``immediately aware what the tools in post will and will not hallucinate,'' while P3 appreciated knowing ``what needs to be AI-generated instead of burning tokens and leave it up to fate.''

Participants rated the \previs{} and slider a median of 6 of 7 for exploring acceptable trade-offs (DG2). When navigating between the \capturedmove{} and \targetmove{}, stopping points reflected qualitative judgments about scene content rather than a numerical threshold (Figure~\ref{fig:study-comfort}B). Participants stopped where performer identity or facial fidelity degraded (P2, P3), where the intended move survived with the actor at low risk (P6), or where, as P5 cautioned, ``you might as well have AI make the whole movie.''

Rather than passively accepting generation costs, participants actively reshaped the input by revising the \targetmove{} (DG4, P5, P6, P7). Both modifying keyframes and adjusting the slider to alter model conditioning, updating the \previs{} interactively. Session logs revealed participants iteratively alternating between both controls: adjusting keyframes, sweeping the slider to inspect updated support, and refining the path again. P7 edited keyframes 25 times to ``limit the visibility of generated wall at the end,'' and P6 made 20 adjustments until the path retained ``the effect I wanted to achieve and there is less hallucination occurred.''

\subsubsection{Selecting Targeted Responses for Distinct Deficits}
Selecting a highlighted deficit in the \previs{} opens a response card with tailored remedy actions (DG3, DG5), letting participants preview solutions before deciding how to respond. For static background gaps, participants could launch a guided pickup or pull back the slider. For unobserved performer angles, where static pickups cannot recover live action, the interface offered retiming proposals, path edits, or replacement takes. With capture available on set in Part~2, participants prioritized filming targeted pickups for static deficits over exhausting the crew on repeated full takes (P1, P2, P3, P4, P6). P3 noted that getting ``pickup shots just for missing info instead of trying to do one perfect shot over and over again'' helps most with ``budgeting and choose the important shots to keep.'' P5 valued knowing ``what wiggle room we'd have on a shot'' on tight schedules, and P6 noted that deciding on set ``is better than regretting a decision later on when discovering a problem in editing.'' Ratings for whether on-set warnings helped decide between a pickup, a retake, and synthesis yielded a median of 6 of 7.

When addressing performance deficits or director requests to alter pacing, participants explored both interactive timing adjustments and automated proposals. Decoupling camera pacing from the spatial path made tuning timing easy (median 6 of 7, with P4, P5, and P6 rating 7 of 7). P6 observed that ``with the timing tool I don't need to follow the plan exactly and still achieve the timing effect I want.'' When evaluating automated retiming proposals that linger in supported views and accelerate through unobserved angles (Appendix~\ref{app:retime}), participants appreciated the proposal as a structural reference, but exercised caution in dynamic scenes. Retiming modifies only camera progress along its path while the recorded performer continues at natural speed on an unchanged clock. Even so, because camera speed guides emotional rhythm and audience attention, automated accelerations risked clashing with the performer's dramatic beats (P4, P5). As P4 noted, for a scene ``where an actor is performing because the performance changes every time,'' preserving dramatic intent took precedence over automated schedules. Participants therefore preferred manually sculpting camera timing, shortening the move spatially via keyframes and the compromise slider, or filming an intentional retake.

No participant was willing to delegate the performer to the model in either Part~1 or Part~2. Across all seven participants, facial fidelity marked the strict boundary where slider exploration stopped. P4 insisted that ``nothing to be hallucinated on the face,'' P7 remained cautious about generating human faces consistently, and P5 emphasized that synthetic infill compromises authentic performance, noting that ``false emotions, and dialogue also would leave me uncomfortable.''

Beyond faces and performance, participants drew boundaries wherever the production had invested creative effort. These included key props (P6), dressed sets (P2), critical story beats (P4), and on-screen text (P3, P6). These decisions reflected narrative importance rather than pixel area. As P1 observed, ``the amount of infill is less concerning than the quality of the regions that need infill.'' Deciding what to leave to the model was recognized as a shared crew decision, with P5 likening real-time diagnostic checks to ``the existence of the script supervisor'' when agreeing on boundaries with the director.

\subsubsection{Filmmaker Comfort Grew with Active Agency and Control} 
Evaluating whether footage is sufficient to move on is an essential step on set. Confidence to wrap a scene knowing how well the take supported the target shot received a median rating of 6 of 7 (Figure~\ref{fig:study-comfort}C). P1 gave a more conservative rating (3 of 7), noting that on-set wrap confidence assumes the target path is already known: ``The biggest hurdle is it relies on knowing some aspects of re-shooting while shooting. If I knew what the desired paths would be, I would have used those to begin with.'' When camera moves are chosen during editing rather than planned on set, P1 valued the feedback for understanding scene limitations during post-production handoffs rather than declaring a scene wrapped.

Across the entire workflow, filmmaker comfort with generative reshooting increased for all participants (Figure~\ref{fig:study-comfort}A), rising from a median of 4 before the study to 6 after experiencing the on-set system. Participants attributed this increase to a sense of agency and transparency rather than uncritical trust in generative models (P2, P3, P4, P5, P6, P7). P5 shifted from viewing generative tools as ``unfaithful to my profession'' to concluding that ``now it feels like a helper.'' P6's comfort grew even as he became ``more cautious about AI generated shots,'' because diagnostic feedback allowed him to actively prevent hallucinations ``by controlling my shot angles.'' Three participants anticipated attempting camera moves they would not have risked before (P1, P2, P7). P7 explained that previsualizing support provides the evidence needed to secure ``buy-in from DP and their department,'' enabling crews to plan ``unusual, ambitious shots and maximizing the likelihood of a positive outcome.''

\section{Discussion}
\label{sec:discussion}

We reflect on how previsualizing \burdenterm{} can change how filmmakers work with generative models, and discuss several design tensions and considerations for generative tools that reach beyond filmmaking.

\subsection{Seeing Hallucination Risk Changes How Filmmakers Shoot}
\label{sec:discussion:shooting}

Filmmaking is an iterative loop: crews plan a shot, rehearse it, roll a take, and repeat until the result is good enough to move on. Moving on is a production call that balances creative goals against time and budget. Under tight schedules, crews often move on and rely on post-production to salvage unresolved defects (P3, P5). Generative reshooting expands what post-production can fix, and with it, what a crew might be willing to leave uncaptured. Yet what the model will actually have to invent remains unknown until the set is struck and it is too late to capture more. The \previs{} brings that check directly into the production loop, before the call to move on.

On set, this visibility relieved the pressure of single-take perfection. In conventional filming, any physical departure from the planned trajectory or timing risks ruining the take. When filmmakers can inspect how departures map onto model support, single-take perfection gives way to capturing coverage. Operators can focus on capturing the live performance, knowing that secondary angles or background surfaces can be supplemented through quick pickups (P1, P2, P3). Rather than surrendering creative control to an automated model, filmmakers use diagnostic feedback to decide where physical capture is necessary and where synthetic infill is acceptable. Comfort grew because filmmakers retained active agency over what the model received, rather than relying on blind trust in generative tools (P5, P6).

We note that not every shot warrants this workflow. For routine setups, traditional in-camera filming and quick retakes remain the fastest approach. Our workflow instead targets complex or physically constrained moves where repeated takes are costly. On set, this is only feasible because the \proxy{} decouples interactive decision-making from generative computation. Running a full video diffusion model live would introduce latency and compute demands that immediately stall an active crew. Instead, evaluating physical coverage in milliseconds on device provides just enough structural feedback for crews to decide whether to revise the move, capture more, or accept synthesis. This architectural separation suggests a broader pattern for creative tools that integrate large generative models, such as 3D environment design, virtual production, and animation. In settings where creators rely on rapid, interactive iteration, providing fast structural proxies allows immediate steering without forcing users to wait on slow model inference.

\subsection{Tension Between Hallucination Risk and Creative Intent}
\label{sec:discussion:tensions}

Generative reshooting spans a spectrum between the challenges of cinematography on one end, and the risks of hallucination on the other. Decoupling camera path from camera timing gives filmmakers a clean way to plan moves along this spectrum, walking the spatial path first and sculpting its rhythm separately. It also enables automated optimization for dynamic subjects (Section~\ref{sec:system:respond}, Appendix~\ref{app:retime}). While static background geometry can always be supplemented with later pickups, a live performance happens only once. Because creators cannot film extra takes to patch a fleeting performance, our space-time search finds schedules that synchronize the authored path with the take, holding when the performer was recorded and hurrying through unobserved angles. Yet camera timing also shapes dramatic pacing and emotional tension, which an automated schedule can easily disrupt. In our study, participants valued having the search surface well-supported timing options, but preferred to sweep the slider and adjust keyframes themselves to protect their dramatic intent (DG5).

Filmmakers navigated this tension directly through the compromise slider. Rather than forcing a binary choice between the recorded take and an ungrounded hallucination, the slider lets creators explore a continuous design space. In post-production, where reshooting is impossible (Part~1), participants used the slider to find a balance where a camera move gained cinematic sweep while ``the footage still held'' (P2). This turned camera selection into an authored trade-off between visual ambition and model support. On set (Part~2), seeing this trade-off in real time changed how participants approached capture. When the \previs{} revealed that an ambitious target move required unseen background coverage, filmmakers did not have to pull the move back to the recorded take. Instead, guided pickups gave them a third option: capturing targeted physical evidence to support the full move they envisioned (DG2, DG4).

This dynamic suggests a transferable design strategy for creative human-AI tools, from generative CAD and animation to synthesized audio. When interactive systems frame AI as an all-or-nothing handoff, users often feel their authorial intent is compromised. By factorizing user intent into separable components, such as isolating spatial path from temporal rhythm, and providing continuous controls between raw input and generative completion, tools can support collaborative steering. Practitioners can then author the expressive core of a work themselves while delegating routine infill to the model.

\subsection{Tension Between Plausibility and Authenticity}
\label{sec:discussion:proxy}

Future improvements in generative models will reduce visual glitches and generation times. Even with higher visual fidelity, the risk of hallucination remains. Photorealistic pixels simply make inventions harder to detect. An advanced model can generate convincing facial expressions and background geometry that appear natural while bearing no relation to the physical scene. This exposes the tension between plausibility and authenticity: balancing visual enhancement against the risk of fabricating details that were never recorded. While filmmakers may accept generative synthesis for secondary background elements, our participants drew a firm line at human performance. For cinematographers, capturing a scene is the joy of the craft, not a passive handoff to post-production. Cinematographers cautioned against generative delegation that displaces the operator's role on set (P5), emphasizing that reshooting must remain ``faithful to the footage that I shot'' (P4). For both practitioners, an authentic take captures deliberate choices made on set that a generative model should not rewrite.

Previsualizing \burdenterm{} establishes a clear boundary of visual responsibility. Showing captured evidence clarifies where a model reconstructs recorded reality and where it invents new content. In our study, participants readily allowed generative infill on textureless background walls, while strictly protecting human faces, emotional expressions, and story-critical props. This selective delegation offers a model for other domains where visual authenticity is critical, such as cultural heritage preservation, scientific visualization, and forensic documentation. In these settings, photorealistic outputs can easily mask synthetic inventions as recorded facts. Visualizing where a model lacks evidentiary support enables domain experts to selectively delegate routine synthesis while protecting the integrity of what was recorded.

\subsection{Limitations and Future Work}
\label{sec:discussion:limitations}

\namedparagraph{Bounded by current model capabilities}
Capture and previsualization run entirely on the mobile device, while final generation occurs offline in post-production. The rendered result is therefore bounded by downstream model constraints. Current video models generate only a few seconds per call due to GPU memory limits, and slow render speeds require creators to iterate on the lightweight proxy. Splitting longer trajectories into chunks creates visual discontinuities at boundaries because current architectures do not share attention across separate generative passes. While our pipeline bridges these seams by regenerating short transitional clips (Appendix~\ref{app:bridge}), doing so requires additional post-processing time.

\namedparagraph{Preview quality and model uncertainty}
The \previs{} visualizes physical support rather than final render appearance. Like editing proxies that enable fast playback before 4K conforming, our point cloud decouples interactive review from slow generative renders. Because point-cloud conditioning is standard across recent reshooting architectures~\cite{lin2026vista4d,zhang2025recapture,cao2026freeorbit4d}, our diagnostic pipeline functions as a model-agnostic, plug-and-play front end. This physical coverage provides a conservative bound for what unobserved angles must hallucinate, but models differ in how effectively they borrow distant frames or synthesize texture. Future work could train lightweight networks to translate our geometric signals into model-specific uncertainty estimates without running heavy diffusion passes on set.

\namedparagraph{Camera form factors and on-device compute}
Live tracking, 3D reconstruction, performer segmentation, and diagnostic evaluation all execute on a single mobile device. While mobile devices provide a convenient all-in-one platform for on-set demonstration, extended sessions cause thermal throttling that degrades processing performance, affecting video segmentation first. Furthermore, professional productions rely on specialized camera packages and external monitoring ecosystems. Future work could decouple these roles across different camera types, pairing handheld mobile directors' monitors with cinema camera rigs or drones, while optimizing on-device pipelines to maintain interactive frame rates under sustained production use.

\section{Conclusion}
\label{sec:conclusion}

We have presented \workflowname{}, a workflow that lets filmmakers see what a generative reshoot would have to hallucinate before anything is generated, and decide on set whether to revise the move, retime, capture more, or accept synthesis. Our key contribution is the \previs{} of \burdenterm{}, rendered from the same \proxy{} the model consumes, which makes that decision cheap enough to revisit after every edit and available wherever a take can be reconstructed. In our study, experienced filmmakers grew more comfortable with generative reshooting once they could evaluate hallucination risk on set and retain active control over what the model receives. Navigating the spectrum between the physical challenges of cinematography and the risks of hallucination will define the future of the craft. We expect shared representations like this \proxy{}, which creators can visually steer and generative models can consume as conditioning, to become essential as practitioners navigate that spectrum in production.

\bibliographystyle{ACM-Reference-Format}
\bibliography{bibliography}

\clearpage

\appendix
\newcommand{\dataflowbody}{%
  \centering
  \includegraphics[width=\textwidth]{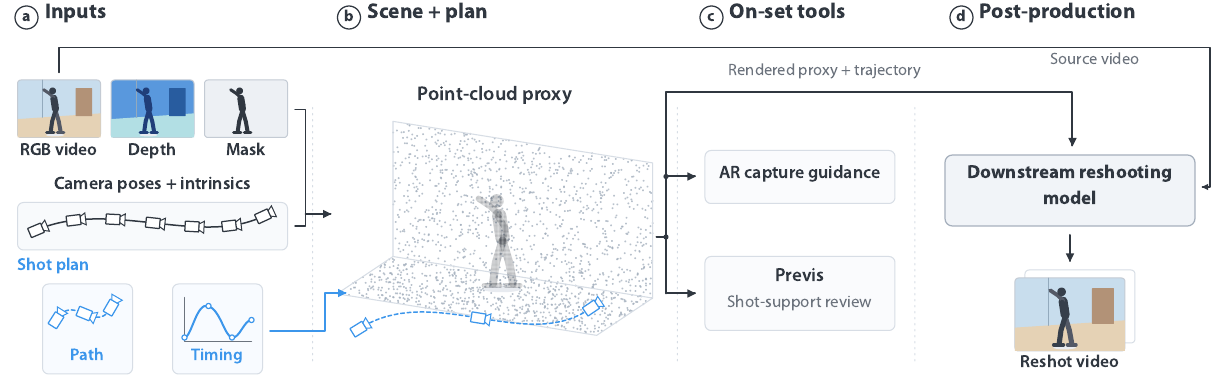}
  \caption{\textbf{Inputs, shared representations, and outputs of the workflow.}
  (a) Video, depth, camera poses and intrinsics, and performer masks provide
  the recorded evidence; path and timing specify the editable shot plan.
  (b) The plan is placed in a point-cloud proxy of the scene and performer.
  (c) This shared scene and plan support AR capture guidance and
  shot-support review before generation.
  (d) After review, the source video, proxy, and selected camera trajectory
  provide the inputs for downstream conditioning and offline reshooting.
  The proxy is rendered along the selected trajectory to prepare
  model-specific conditioning views and masks. All illustrations are schematic.}
  \label{fig:dataflow}
  \Description{A compact data-flow diagram with four columns.
  Inputs show RGB, depth, and performer-mask views, camera poses and intrinsics,
  and two square panels for path and timing. These feed a dotted room and
  performer with the planned camera path. A shared connector branches to
  boxes for AR capture guidance and previs with shot-support review,
  and to a downstream reshooting model in a separate post-production column.
  The source RGB video also connects directly to the model, without
  repeating the inputs. The model outputs a reshot video.}}
\makeatletter
\if@twocolumn
  \twocolumn[{\def\@captype{figure}\dataflowbody\vspace{1.5em}}]
\else
  \begin{figure*}[t]
    \dataflowbody
  \end{figure*}
\fi
\makeatother

\section{Inputs and Outputs of the Workflow}
\label{app:dataflow}

Figure~\ref{fig:dataflow} summarizes how the prototype connects recorded
evidence, the editable shot plan, and downstream generation. On set, the
mobile device records camera poses via visual-inertial odometry, scene depth with hardware LiDAR, and performer masks via EdgeTAM~\cite{zhou2025edgetam}. For ordinary video, depth, camera poses, and performer masks are estimated offline with $\pi^3$~\cite{wang2026pi} (or Depth Anything~3~\cite{lin2025depth}) and SAM~3~\cite{carion2026sam} (Section~\ref{sec:system:offline}). Both routes support the same review of
moves between the captured and target moves before the selected move is passed to offline reshooting.

\section{Path Representation}
\label{app:splines}
The path must pass through every keyframe the filmmaker placed, since each one is a framing they stood in and approved, and it must be smooth enough that following it produces no jerky guidance. We therefore interpolate rather than approximate, and choose splines that are continuous in velocity. Position uses a centripetal Catmull-Rom spline~\cite{yuksel2011parameterization}, which passes through every keyframe and, unlike the uniform variant, does not overshoot or loop when two keyframes are close together. The first and last segments need a tangent from a neighbor that does not exist, so we reflect the adjacent keyframe across each end to supply one. Orientation is interpolated separately, because rotations do not live in the same space as positions and averaging them directly produces wobble; we use squad~\cite{shoemake1987quaternion}, the spherical counterpart of Catmull-Rom, which turns the camera smoothly through each keyframe's orientation with tangents adjusted for uneven keyframe spacing. A shot-type preset adds interior keyframes to this same fit about a pivot chosen when it is declared (a tapped point, else the subject, else the viewfinder raycast), so a preset and a free stretch are the same kind of thing to the path. Progress along the path only increases with output time, so a retiming can hold or slow the camera but never send it backward.

\section{Bridging Chunk Seams}
\label{app:bridge}

We use Vista4D~\cite{lin2026vista4d} as the downstream model because it was the most recent reshooting model with open weights at the time and conditions on exactly what the workflow exports, the source video and a point cloud rendered from the target camera. It generates 49 frames per call, so a 192-frame move is generated as four chunks, and because each chunk fills in the unobserved scene differently, each join shows as a jump cut. We generate one more clip of native length centered on the join, held to the existing video at both ends and free in the middle, so the join falls inside a single generation. The ends are held by nudging the clip toward the existing video at every step, strongly at the edges and not at all near the join, in the manner of RePaint-style inpainting. The result replaces the frames around the join directly, with no blending, no training, and one extra generation per join.

\section{Searching over Timing and Extent}
\label{app:retime}

\begin{figure*}
  \centering
  \includegraphics[width=\textwidth]{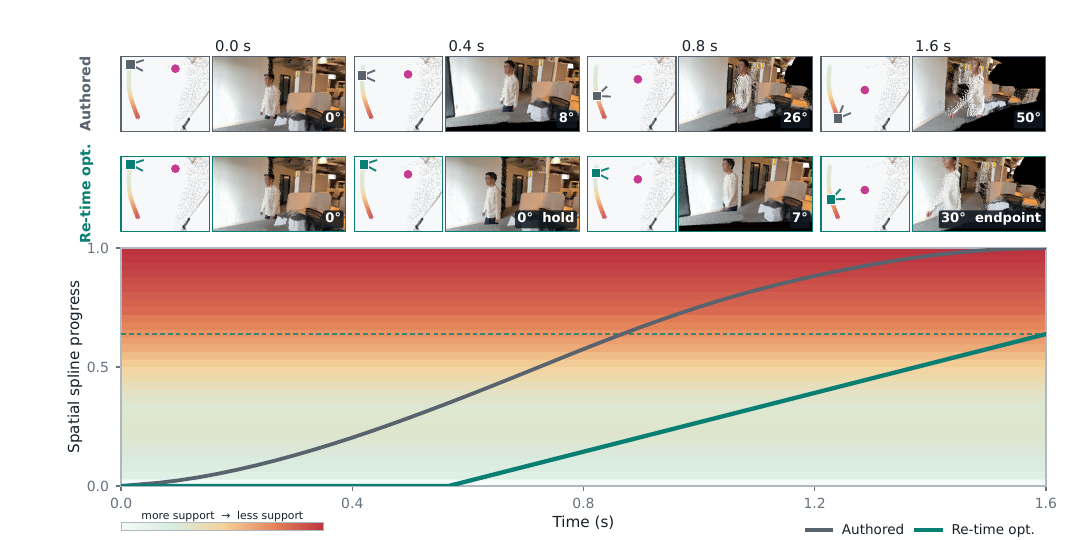}
    \caption{\textbf{Space--time search for a better-supported move.} The grid plots path progress against output time, with the background running from better-supported (bottom) to less-supported (top) views. The authored schedule (gray) reaches the unsupported end of the path and a $50^\circ$ view. The proposal (teal) begins with a hold, advances more slowly to linger in supported views, and stops at a $30^\circ$ endpoint, lowering \burdenterm{}. Equal-time samples above show the views requested by each schedule.}
  \label{fig:retime_optimization}
\end{figure*}

Because timing is authored separately (Section~\ref{sec:system:plan}), the
system can search over it. The \emph{Retime} proposal treats retiming as
synchronizing the authored path with the recorded take. On a grid of path
progress against take time, a schedule is a curve that only moves forward
in both, a hold is a horizontal run, and each cell costs by how far the
authored camera is from where the real camera was at that instant, how
under-resolved the room is from there, and how much of the performer that
view fails to see. Dynamic programming returns the cheapest schedule, with
holds and hand-authored spans frozen so a search cannot undo a deliberate
decision. The search may also stop short of the path's end, and a combined
search finds the least shortening that leaves the performer supported
throughout. Either result is shown as a proposal with its consequences and
applied only if the filmmaker accepts it (DG5), and nothing is offered when
any instant would remain unsupported or unknown.
Figure~\ref{fig:retime_optimization} shows a combined proposal that holds
in a supported view, then advances through the lower-cost part of the grid
to a reduced endpoint.

\begin{figure*}[t]
  \centering
  \includegraphics[width=\textwidth]{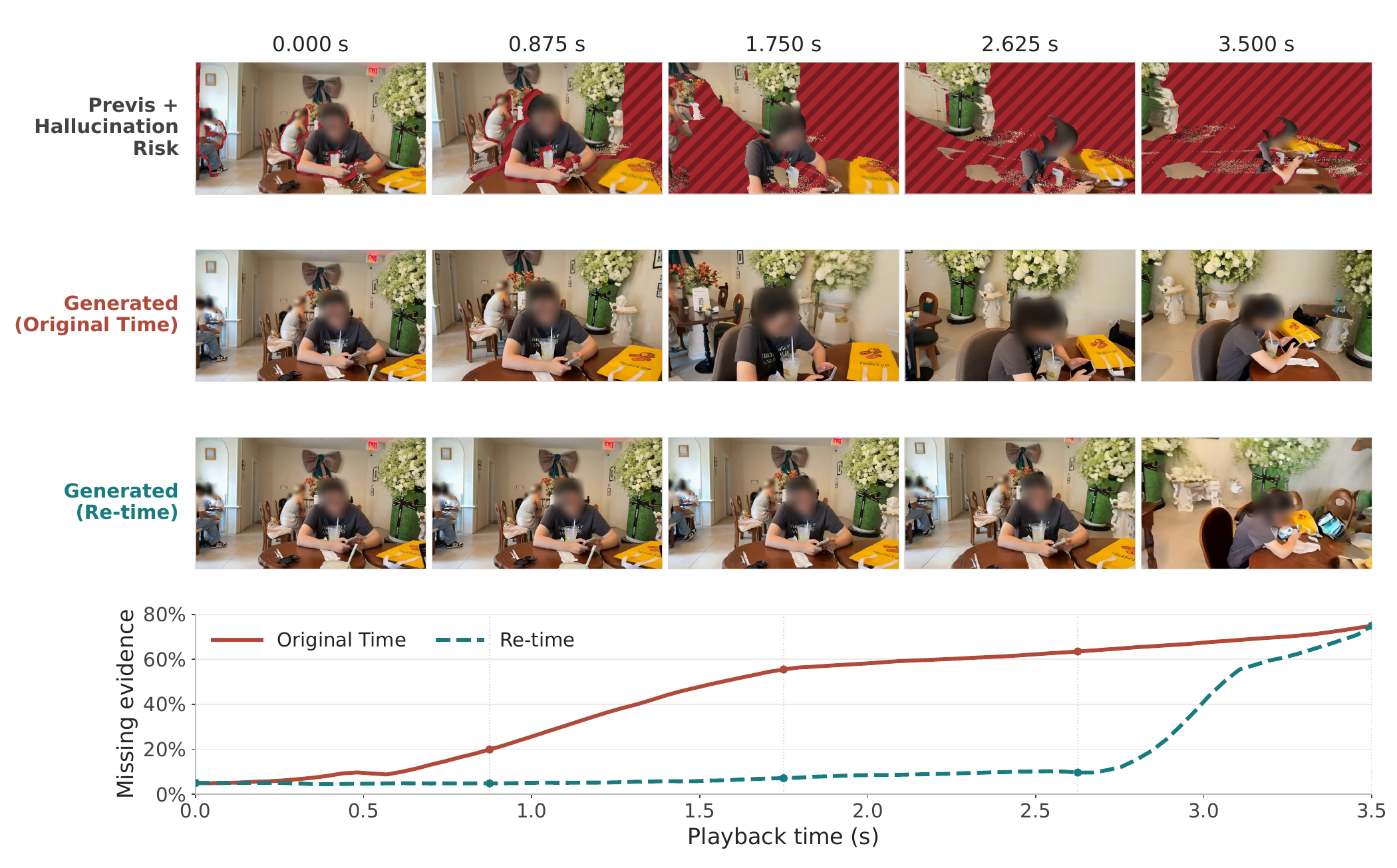}
  \caption{\textbf{Original and retimed reshoots at matched playback times.}
  Previs reveals substantial missing coverage along the target path.
  Original timing prolongs views the generator must hallucinate; retiming
  preserves the full path and 3.5-second duration, lingering where evidence
  is stronger and whip-panning through poorly supported views to reduce
  hallucination. The plot shows the fraction of projected point-cloud pixels
  without evidence, not measured hallucination error. Faces are blurred for
  anonymity. Video is provided in the supplemental material.}
  \label{fig:retime_results}
  \Description{Three rows compare original-timing previs, original-timing
  generation, and retimed generation at five matched playback times from
  0 to 3.5 seconds. Red hatching in the previs marks missing coverage.
  The original generation progressively reveals unsupported parts of the
  room, while the retimed generation stays near the supported starting view
  before moving rapidly to the endpoint. Below, a line plot shows missing
  evidence rising throughout the original schedule but remaining low for
  most of the retimed schedule before rising sharply near the end.}
\end{figure*}

Figure~\ref{fig:retime_results} compares generated reshoots with original
and revised timing along the same full camera path.

\section{Evaluation Materials and Individual Responses}
\label{app:study}

This appendix documents the study responses summarized in
Section~\ref{sec:evaluation}. Part~1 comprised an introductory sequence and a
second scene in an interactive browser viewer. In Part~2, participants used
the mobile AR application that demonstrates \workflowname{} and answered
questions about their experience and its use in future productions.
The two parts and questionnaire sections followed a fixed order.

\subsection{Part 1 Tasks and Scoring}

The initial pin task paired the source take with a generated image at frame
42. Participants were told that the source camera had filmed the actor from
the side and asked to click content the camera had never seen. The reveal
showed the reference mask and a separately annotated face region. A coverage
lesson then compared the reconstructed proxy with a generated result, followed
by a slider demonstration. This was an instructed exercise, not an unaided
test of detecting generated content.

The second scene used the orbit target at frame 86 in the recorded sessions.
Participants first pinned missing content in the proxy with overlays hidden,
then viewed coverage and performer warnings, chose a slider stop, completed
two checks, and explored the viewer while answering closing questions. The
resolution overlay was disabled. The checks assessed interpretation of the performer
heat scale and comparison of generation needs between two camera stops.

Each participant started the acceptance task at $\lambda=0.5$, and the accepted
stop is the submitted response.

Participants explained their chosen stop with a prompt suggesting
actor-face versus background-room tradeoffs. Closing items
asked about comfort with generative video reshooting,
preferred preview timing (on set, editing, both, or neither), reasons for those
answers, and one desired change. P1 and P4 asked to edit the camera path directly, P2 wanted a limit on
viewpoint departure, P3 asked for a preview of a paused frame, P5 for a
preview of the generated options, P6 for an automatic pan that surveys
the scene while shooting, and P7 for integration with editing software such
as Adobe Premiere. These requests are individual suggestions, not
shared themes.

The initial and post-exercise items assessed comfort with generative video
reshooting on the same seven-point scale, from 1 (not at all comfortable)
to 7 (completely comfortable).

\subsection{Part 2 Ratings}

The Part~2 questionnaire asked participants to reflect on their use of the
mobile AR application. Questions covered planning, guidance, review, and future workflow
integration in that order. The questionnaire included interface illustrations
and examples of moves difficult or expensive to execute physically, such as
robotic camera movement and dolly crews. Table~\ref{tab:study-part2}
summarizes all Part~2 ratings. Agreement items used 1 (strongly disagree), 4 (neutral),
and 7 (strongly agree). The comfort item used 1 (not at all comfortable) and
7 (completely comfortable). Item numbers in the table match the questionnaire.

The storytelling question additionally suggested actors, key props, faces, and
emotions versus background geometry, ceilings, and distant walls. The guidance
and review questions also described expected benefits. Accordingly, the
responses should be read in the context of these examples and framings.

\begin{table*}[t]
\centering
\small
\caption{All Part~2 ratings after using the mobile AR application, with
abbreviated labels for the questionnaire items.
The final column gives the median and observed minimum--maximum.}
\label{tab:study-part2}
\begin{tabular}{@{}llcccccccc@{}}
\toprule
\textbf{Item} & \textbf{Statement topic} & \textbf{P1} & \textbf{P2} & \textbf{P3} & \textbf{P4} & \textbf{P5} & \textbf{P6} & \textbf{P7} & \textbf{Median (min--max)} \\
\midrule
\multicolumn{10}{@{}l}{\textit{Planning and capture}} \\
1.1 & Easy to plan by walking, saving viewpoints & 7 & 7 & 7 & 6 & 7 & 6 & 5 & 7 (5--7) \\
1.2 & Easy to tune timing apart from the path & 5 & 5 & 5 & 7 & 7 & 7 & 6 & 6 (5--7) \\
2.1 & Guidance freed attention for the live scene & 6 & 6 & 4 & 6 & 7 & 6 & 5 & 6 (4--7) \\
\midrule
\multicolumn{10}{@{}l}{\textit{Review and decisions}} \\
3.1 & Clearly saw where and why AI must invent & 7 & 7 & 7 & 7 & 7 & 7 & 5 & 7 (5--7) \\
3.2 & Slider helped explore acceptable trade-offs & 7 & 6 & 7 & 5 & 7 & 6 & 6 & 6 (5--7) \\
3.3 & Warnings helped choose pickup, retake, or AI & 6 & 6 & 7 & 6 & 7 & 7 & 5 & 6 (5--7) \\
\midrule
\multicolumn{10}{@{}l}{\textit{Overall workflow}} \\
4.1 & Comfort using reshooting in editing & 6 & 6 & 6 & 6 & 5 & 6 & 7 & 6 (5--7) \\
4.3 & Felt in control of creative decisions & 6 & 5 & 6 & 5 & 5 & 6 & 5 & 5 (5--6) \\
4.4 & On-set feedback worth the extra time & 5 & 6 & 6 & 5 & 5 & 7 & 6 & 6 (5--7) \\
4.5 & Confident wrapping, knowing footage support & 3 & 5 & 7 & 6 & 6 & 6 & 6 & 6 (3--7) \\
\bottomrule
\end{tabular}
\end{table*}

\end{document}